\documentstyle[twocolumn,floats,epsf,ifthen,aps,prb]{revtex} 

\begin{document} 
 
\draft 
%\preprint{draft} 

\title{ 
Continuous quantum measurement of a double dot}

\author{Alexander N. Korotkov}
\address{ 
Department of Physics, State University of New York at Stony Brook, 
Stony Brook, NY 11794-3800 \\ 
and \\
GPEC, Departement de Physique, Facult\'e des Sciences de Luminy, 
Universit\'e de la M\'editerran\'ee, 
13288 Marseille, France 
} 
\date{\today} 
 
\maketitle 
 
\begin{abstract} 
        We consider the continuous measurement of a double 
quantum dot by a weakly coupled detector (tunnel point contact 
nearby). While the conventional approach describes the gradual 
system decoherence  due to the measurement, 
we study the situation when the detector output is explicitly 
recorded that leads to the opposite effect: gradual purification 
of the double-dot density matrix.  
Nonlinear Langevin equation is derived for the random 
evolution of the density matrix which is reflected and caused 
by the stochastic 
detector output. Gradual collapse, gradual purification,  
and quantum Zeno effect are naturally described by the equation.
We also discuss the possible experiments to confirm the theory.
\end{abstract} 
 
%\pacs{PACS numbers: 73.23.-b, 03.65.Bz}
 
\narrowtext 
 
%\vspace{1ex} 

        The problem of quantum measurements has a long history,
however, it still attracts considerable attention and even causes
some controversy, mainly concerning the wavefunction 
``collapse'' (see, e.g., \cite{Wheeler,Braginsky}).  
        Among various modern approaches to this problem let us 
mention the idea of replacing the collapse postulate by the gradual 
decoherence of the density matrix due to the interaction with the 
detector \cite{Zurek} and 
the approach of a stochastic evolution of the wavefunction
(see, e.g., 
\cite{Gisin,Carmichael,Plenio,Mensky,Presilla,Gagen,Hegerfeldt,Dalibard}). 
The latter approach (which is used in the present paper) can
describe the selective measurements for which the system
evolution is conditioned on the particular measurement result
(other keywords of the approach are: quantum trajectories, quantum 
state diffusion, quantum jumps, etc.). 
The renewed interest in the measurement problem 
is justified by the development of experimental technique,
which allows more and more experimental studies of quantum
measurement in optics and mesoscopic structures. 
\cite{Itano,Aspect,Brune,Buks,Tittel,Nakamura}  
The problem also has a close connection to the rapidly growing 
fields of quantum cryptography 
and quantum computing.\cite{q-comp} 

        In the recent experiment \cite{Buks} with  ``which-path''
interferometer the suppression of Aharonov-Bohm interference 
due to the detection of which path an electron chooses, was observed. 
The weakly coupled quantum point contact was used as a detector.
The interference suppression in this experiment can be quantitatively 
described by the decoherence due to the measurement process. 
\cite{Gurvitz,Aleiner,Levinson,Stodolsky}   

We will consider a somewhat different setup: 
two quantum dots occupied by one electron and a weakly coupled 
detector (point contact nearby) measuring the position of the electron.
The decoherence of the double-dot density matrix  due to continuous 
measurement in this setup has been analyzed  in Refs.\ 
\cite{Gurvitz,Stodolsky}. However, the decoherence approach
cannot describe the detector output that is a separate 
problem analyzed in the present paper. 
        We answer two interrelated questions: 
how the detector current behaves in time 
and what is the proper double-dot density matrix 
for a particular detector output. 
We show that the models of point contact considered in Refs.
\cite{Gurvitz,Aleiner,Levinson} describe an ideal detector.  
In this case the density matrix decoherence is just a 
consequence of averaging over all possible 
measurement results. For any particular detector output 
our equations allow the evolution of pure wavefunction
to be followed.
Moreover, a mixed density matrix can be gradually purified 
in the course of a continuous measurement. 
  
        Similar to Ref.\ \cite{Gurvitz} let us describe
the double-dot system and the measuring point contact by the Hamiltonian
       $ {\cal H}={\cal H}_{DD}+{\cal H}_{PC}+{\cal H}_{int}$,
where 
${\cal H}_{DD} = (\varepsilon /2) (c_1^\dagger c_1-c_2^\dagger c_2)+ 
H (c_1^\dagger c_2+ c_2^\dagger c_1) $
is the Hamiltonian of the double-dot,   
${\cal H}_{PC}=\sum_l E_la_l^\dagger a_l +\sum_r E_r a_r^\dagger a_r +
\sum_{l,r} T (a_r^\dagger a_l+a_l^\dagger a_r) $
describes the tunneling through the point contact ($T$ and $H$ are real), 
and ${\cal H}_{int}= \sum_{l,r} \Delta T \, c_2^\dagger c_2 
(a_r^\dagger a_l + a_l^\dagger a_r)$,
i.e.\ the tunneling matrix element for the point contact is $T$ or 
$T+\Delta T$ depending on which dot is occupied.
So, the average current $I_1=2\pi T^2 \rho_l \rho_r e^2V/\hbar$ flows 
through the detector when the electron is in the first dot
($V$ is voltage across the tunnel contact, $\rho_l$ and $\rho_r$ are 
the densities of states), while the current is
$I_2=I_1+\Delta I=2\pi (T+\Delta T)^2\rho_l\rho_r e^2V/\hbar$ 
when the second dot is occupied.

        We make an important assumption of weak coupling between
the double-dot and the detector (a better term would be the
``weakly responding'' detector),
        \begin{equation}
| \Delta I | \ll I_0= (I_1 +I_2)/2,
        \label{weak}\end{equation}
so that many electrons, $N \gtrsim (I_0/\Delta I)^2 \gg 1$, should 
pass through the point contact before
one can distinguish which dot is occupied. 
This assumption allows the classical description 
of the detector, namely, to neglect the coherence between 
the quantum states with different number of electrons passed 
through the detector. \cite{bigDI} 

        The decoherence rate $\Gamma_d = (\sqrt{I_1/e}-\sqrt{I_2/e})^2/2$
of the double-dot density matrix ${\bf \sigma} (t)$ due to 
the measurement by tunnel point 
contact has been calculated in Ref.\ \cite{Gurvitz}.
In the weakly-responding limit (\ref{weak}) it can be replaced
by $\Gamma_d= (\Delta I)^2/8eI_0$ or by the expression 
        \begin{equation}
\Gamma_d = (\Delta I)^2/4S_I,
        \label{gam_d-m}\end{equation} 
where $S_I=2eI_0$ is the usual Schottky formula for the detector 
shot noise spectral density $S_I$. 
Equation (\ref{gam_d-m}) has also been obtained in Refs.\
\cite{Aleiner,Levinson,Stodolsky} for the quantum point contact as a detector,
the difference in that case is $S_I=2eI_0 (1-{\cal T})$ 
where ${\cal T}$ is the transparency of the channel \cite{Lesovik} 
(while above we implicitly assumed  
${\cal T}\ll 1$). \cite{largeT,phase} 
        Notice that the decoherence rate (\ref{gam_d-m}) was derived 
in Refs.\ \cite{Gurvitz,Aleiner,Levinson,Stodolsky} without any account of
the information provided by the detector, implicitly assuming that
the measurement result is just ignored. Now let us study how
this additional information affects the double-dot density matrix.

\vspace{0.2cm}

        We start with the completely classical case in which there 
is no tunneling between dots ($H=0$) and
the initial density matrix of the system does not have nondiagonal 
elements, $\sigma_{12}(0)=\sigma_{12}(t)=0$. 
We can assume that the electron is actually located in one of the 
dots, but it is not known in which one, and that is why we 
use probabilities $\sigma_{11}(0)$ and $\sigma_{22}(0)=1-\sigma_{11}(0)$. 
The detector output is the fluctuating current $I(t)$. 
The fluctuations grow when $I(t)$ is examined 
at smaller time scales, so some averaging in time (``low-pass
filtering'') is necessary, at least in order to neglect the problem 
of individual electrons passing through the point contact. 
Let us always work at sufficiently low 
frequencies, $f\sim \tau^{-1}\ll S_I/e^2$, for which the possible 
frequency dependence of $S_I$ can be neglected. 

        The probability $P$ 
to have a particular value for the current averaged over time $\tau$,  
$\langle I \rangle =\tau^{-1} \int_0^\tau I(t) dt$, 
 is given by the distribution 
        \begin{eqnarray}
P(\langle I\rangle , \tau)=\sigma_{11}(0) \, P_1 (\langle I\rangle ,\tau )+ 
\sigma_{22}(0) \, P_2(\langle I\rangle ,\tau ), 
        \label{prob} \\
P_i(\langle I\rangle , \tau ) = (2\pi D)^{-1/2} 
\exp \left[ -(\langle I\rangle -I_i)^2/2D \right] , 
        \label{Gauss}\end{eqnarray}
where $ D = S_I/2\tau$.
Notice that these equations obviously do not change if we divide the 
time interval $\tau$ into pieces and integrate over all 
possible average currents for each piece 
(to consider only positive currents, the typical 
timescale $\tau$ should be sufficiently long, $S_I/\tau \ll I^2$, 
that is always satisfied within the assumed low frequency range).

After the measurement during
time $\tau$ we acquire additional knowledge about the system 
and should change the probabilities $\sigma_{ii}$ according to the 
standard Bayes formula \cite{Bayes} for a posteriori probability
(taking into account particular detector result $\langle I\rangle $:
        \begin{eqnarray}
\sigma_{11} (\tau ) &&= \sigma_{11}(0) \exp [ 
-(\langle I\rangle -I_1)^2/2D ] 
        \nonumber \\
&&\, \times \left\{ \sigma_{11}(0)\exp [ -(\langle 
 I\rangle -I_1)^2/2D ] \right.  
        \nonumber \\
 && \,\,\,\,\,\,\,\, + \left. \sigma_{22}(0) 
\exp  [- (\langle I\rangle -I_2)^2/2D] 
\right\} ^{-1} ,
        \nonumber \\
\sigma_{22} (\tau ) &&= 1-\sigma_{11} (\tau ) .
        \label{s1-2}\end{eqnarray}

	Notice that we have considered so far the purely classical
measurement and did not use any ``collapse'' postulate. Nevertheless,
Eq.\ (\ref{s1-2}) can be interpreted as a gradual ``localization''
of the electron in one of the dots due to acquired information.

\vspace{0.2cm}

        Now let us assume that the initial state is fully coherent,
$|\sigma_{12}(0)|=\sqrt{\sigma_{11}(0)\sigma_{22}(0)}$, while still 
$H=0$. Since the detector is sensitive only 
to the position of an electron, the detector current 
will behave {\it exactly} the same way as in the case above. 
So, after  the  measurement  during  time  $\tau$   we   should 
assign the same values for $\sigma_{ii}(\tau )$  
as in Eq.\ (\ref{s1-2}), but the question is not so trivial
for the nondiagonal matrix element $\sigma_{12}(\tau )$.
Nevertheless, we can write the upper bound:
        \begin{equation}
        |\sigma_{12}(\tau )| \leq 
\sqrt{ \sigma_{11}(\tau ) \sigma_{22}(\tau )}. 
        \label{u-b}\end{equation}

        If the actual measurement result is disregarded, 
then the upper bound for $|\sigma_{12}|$ can be calculated 
using the probability distribution of different outcomes given by
Eq.\ (\ref{prob}) and the upper bound (\ref{u-b}) for each
realization: 
        \begin{eqnarray}
| \langle \sigma_{12}(\tau )\rangle _r | \leq 
\int \sqrt{\sigma_{11}(\tau )\sigma_{22}(\tau )} \, 
        P(\langle I\rangle ,\tau ) \, d\langle I\rangle 
        \nonumber \\
= \sqrt{\sigma_{11}(0)\sigma_{22}(0)} \, 
\exp [ -(\Delta I)^{2}\tau /4S_I ] 
        \end{eqnarray}
(here $\langle \,\,\rangle _r$ means averaging over realizations).
This upper bound exactly coincides \cite{strong} with the result 
given by the decoherence approach (\ref{gam_d-m}). This fact
forces us to accept the somewhat surprising statement
that Eq.\ (\ref{u-b}) gives not only the upper bound, but 
the true value of the nondiagonal 
matrix element, i.e.\ the pure state {\it remains pure}   
(no decoherence occurs) during each particular measurement. 
(Actually, this is the usual statement for 
selective measurements, 
\cite{Carmichael,Plenio,Mensky,Presilla,Gagen,Hegerfeldt,Dalibard}
i.e.\ when the detector output is taken into account.) 
        
        Simultaneously, we proved that the point contact detector 
considered theoretically in Refs.\ \cite{Gurvitz,Aleiner,Levinson} 
(the model is confirmed experimentally \cite{Buks}) causes the slowest 
possible decoherence of the measured system, 
 and hence represents an ideal detector in this sense. 
In contrast, the result 
of Ref.\ \cite{Shnirman} shows that a single-electron transistor
\cite{Likh-87}  biased by relatively large voltage 
is not an ideal detector (the non-ideal detector has also been
considered in  Ref.\ \cite{Stodolsky}). Notice, however, that 
in the range of elastic cotunneling \cite{Av-Naz} 
the operation of the single-electron transistor is almost 
equivalent\cite{Shnirman} to the case considered above, and, hence,
it becomes an ideal detector. 

        If the initial state of the double-dot is not purely coherent, 
$|\sigma_{12}(0)|<\sqrt{\sigma_{11}(0)\sigma_{22}(0)}$, 
it can be treated as the statistical combination of purely 
coherent and purely incoherent states with the same
$\sigma_{11}(0)$ and $\sigma_{22}(0)$, then 
        \begin{equation}
\sigma_{12}(\tau )= \sigma_{12}(0) \, \exp 
\left( \frac{i\varepsilon\tau}{\hbar}\right)  
\left[\frac{ \sigma_{11}(\tau ) \sigma_{22}(\tau ) }
 {\sigma_{11}(0) \sigma_{22}(0)}\right] ^{1/2} . 
        \label{s_12-g}\end{equation}

        Equations (\ref{s1-2}) and (\ref{s_12-g})  are the central 
result of the present paper; they give the density matrix 
of the measured system (in the case $H=0$) with account of the 
measurement result.\cite{alternat} 

        These equations can be also used to simulate the detector 
output $I(t)$ and the corresponding 
evolution of the density matrix. For example, in the 
Monte-Carlo method we should first choose
the timestep $\tau$ satisfying inequalities $e^2/S_I \ll \tau
\ll S_I/(\Delta I)^2$ and draw a random number for 
$\langle  I\rangle$ according to the distribution (\ref{prob}). 
Then we update $\sigma_{11}(t)$ and $\sigma_{22}(t)$ using
this value of $\langle I\rangle$ and repeat the procedure many
times [the distribution for the current averaged over 
the interval $\Delta t=\tau$ 
is new every timestep because of changing $\sigma_{ii}(t)$ which
are used in Eq.\ (\ref{prob})].
The nondiagonal matrix element can be calculated at any time
with Eq.\ (\ref{s_12-g}). 

        Using Eqs.\ (\ref{prob})--(\ref{s1-2}), this Monte-Carlo 
procedure can be easily reduced to the following 
nonlinear Langevin-type equation  
(equation for $\sigma_{11}$ is sufficient):
        \begin{eqnarray}
{\dot \sigma}_{11}={\cal R}&&,  \,\,\, \, {\cal R} =
-\sigma_{11}\sigma_{22}\, \frac{2\Delta I}{S_I} 
\left[ I(t)-I_0 \right] 
        \label{Rnew} \\ 
&&= -\sigma_{11}\sigma_{22}\, \frac{2\Delta I}{S_I}
\left[\frac{\sigma_{22}-\sigma_{11}}{2}\, \Delta I +\xi (t)  \right] ,
        \label{s_11-evol}\end{eqnarray}
where the random process $\xi (t)$ has zero average and ``white''
spectral density $S_\xi = S_I$.
        The second expression for $\cal R$ allows the measurement 
to be simulated while the first one can be used to calculate the 
density matrix for given $I(t)$ [in case $H=0$ it can more easily 
be done using Eq.\ (\ref{s1-2})].

\begin{figure}
\centerline{
\epsfxsize=3.0in
\epsfbox{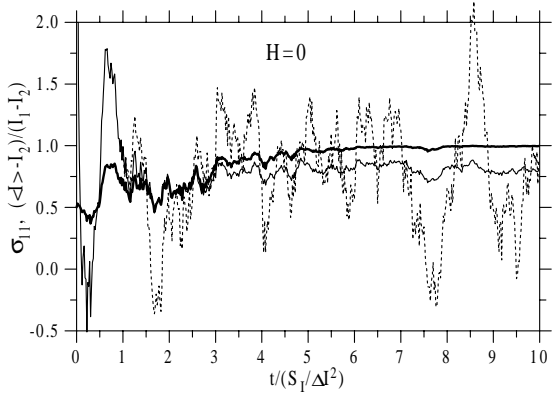}} 
\caption{Thick line:  particular Monte-Carlo realization 
of $\sigma_{11}$ evolution in time during the measurement of uncoupled
dots, $H=0$. 
Initial state is symmetric, $\sigma_{11}(0)=\sigma_{22}(0)=1/2$,
while the measurement leads to gradual localization. 
Initially pure wavefunction remains pure at any time $t$. 
Thin line shows the corresponding detector current 
$\langle I \rangle$ averaged over the whole time interval starting 
from $t=0$ while the dashed line is the current averaged over
the running window with duration $S_I/(\Delta I)^2$.
 }
\label{fig1}
\end{figure}

        Figure \ref{fig1} shows a particular result of the Monte-Carlo
simulation for the symmetric initial state, $\sigma_{11}(0)=
\sigma_{22}(0)=1/2$. The thick line shows the random evolution 
of $\sigma_{11}(t)$. Equation (\ref{s_11-evol}) describes the gradual 
localization in one of the dots (first dot in case of Fig.\ \ref{fig1}).
Let us define the typical localization time as
$\tau_{loc}=2S_I/(\Delta I)^2 $ (we choose the exponential factor
at $\sigma_{11}=\sigma_{22}=1/2$). Then it is exactly equal to the time
$\tau_{dis}=2S_I/(\Delta I)^2$ necessary to distinguish
between two states (defined as the shift of two Gaussians 
(\ref{Gauss}) from $I_0$ by one 
standard deviation), and $\tau_{loc}=\tau_d/2$ where $\tau_d=\Gamma_d^{-1}$.
It is easy to prove that the probability of final localization 
in the first dot is equal to $\sigma_{11}(0)$,   
because $\sigma_{ii}(\tau )$ averaged over realizations is conserved
(the deterministic flow of $\sigma_{11}$ due to the first term 
in square brackets of Eq.\ (\ref{s_11-evol}) is exactly canceled 
on average by the dependence of the diffusion coefficient on 
$\sigma_{11}$).

        The detector current $I(t)$ basically follows the evolution of
$\sigma_{ii}(t)$ but also contains the noise which
depends on the bandwidth. The dashed line in Fig.\ 
\ref{fig1} shows the current
$\langle I(t,t-\Delta t)\rangle =\Delta t^{-1} \int_{t-\Delta t}^t I(t)\, dt$
averaged over the ``running window''
with duration $\Delta t= S_I/(\Delta I)^2$,
while the thin solid 
line is the current $\langle I(t,0)\rangle$ averaged starting from $t=0$. 

\vspace{0.2cm}

        Now let us consider the general case of the double-dot
with non-zero tunneling $H$. If the frequency $\Omega$ of ``internal''
oscillations is sufficiently low, 
$ \Omega = (4H^2+\varepsilon^2)^{1/2}/\hbar \ll S_I/e^2 $,
we can use the same formalism just adding the  
evolution due to finite $H$ (the product $\Omega \tau_{loc}$
is {\it arbitrary}). 
A particular realization can be
either simulated by Monte-Carlo procedure similar to that outlined 
above [now update of $\sigma_{12}(t)$ using Eq.\ (\ref{s_12-g}) should
be necessarily done at each timestep, together with the evolution 
due to finite $H$] 
 or equivalently described by the coupled
Langevin equations 
        \begin{eqnarray}
{\dot \sigma}_{11} &=& -{\dot \sigma}_{22}= (-2H/\hbar) \, 
        \mbox{Im}(\sigma_{12}) +{\cal R},
        \label{s11-g}   \\
{\dot \sigma}_{12} &=& \frac{i\varepsilon }{\hbar} \sigma_{12} 
+\frac{iH}{\hbar}(\sigma_{11}-\sigma_{22}) +
\frac{\sigma_{22}-\sigma_{11}}{2\sigma_{11}\sigma_{22}}\, {\cal R} \sigma_{12} 
        \nonumber \\
&& -\gamma_d\sigma_{12},
        \label{s12-g}\end{eqnarray}
where $\gamma_d =0$ for an ideal detector (see below). 
        The alternative ``microscopic'' derivation of these 
equations can be done for the particular model of Ref.\ \cite{Gurvitz}
and will be presented elsewhere. 

        Notice that in Eqs.\ (\ref{Rnew})--(\ref{s12-g}) the 
derivative is defined as ${\dot \sigma (t)}=\mbox{lim}_{\tau \rightarrow 0}
 [\sigma (t+\tau /2)-\sigma (t-\tau /2)]/\tau$ (Stratonovich formulation  
of the stochastic equations\cite{Oksendal}). The equations would be 
different if the definition 
$[\sigma (t+\tau ) -\sigma (t)]/\tau$  was used (It\^o formulation). 
We use the former one because it gives the correct limit when the noise 
term $\xi (t)$ is replaced by a sequence of smooth functions 
\cite{Oksendal} and 
also because the equations in Stratonovich formalism are physically
more transparent since they do not contain extra terms arising due to 
${\cal R}^2 dt=\mbox{const}$ (for example, the usual calculus rule 
$(fg)'=f'g+fg'$ 
is still valid). To translate Eqs.\ (\ref{Rnew})--(\ref{s12-g})
into It\^o formalism, one would need to add the terms\cite{Oksendal} 
$(S_I/2)F(dF/d\sigma)/2$ where $F$ is the factor before $\xi (t)$. 
This would lead to extra terms $-(\sigma_{22}-\sigma_{11})\Delta I/2$
in square brackets of Eqs.\ (\ref{Rnew})--(\ref{s_11-evol}) and 
extra term $-\sigma_{12}(\Delta I)^2/4S_I$ in Eq.\ (\ref{s12-g}).
Notice that in It\^o formalism the equations become linear (except
for the terms proportional to $\xi (t)$). 

        The simplest way to avoid the possible confusion between 
two formulations of stochastic equations is to use the explicit 
calculation procedures (for finite
$\tau$) described above. However, the difference should be taken 
into account when results of other approaches to the 
stochastic wavefunction evolution 
\cite{Gisin,Carmichael,Plenio,Mensky,Presilla,Gagen,Hegerfeldt,Dalibard} 
are compared. 
For example, this explains the apparent difference between Eqs.\
(\ref{Rnew})--(\ref{s12-g}) and the results of Ref.\ 
\cite{Gagen} for a two-level system (with $\varepsilon=0$ and $\gamma_d=0$) 
derived in a different way. Among various approaches to 
selective quantum measurements, our approach is most 
closely related to the method of restricted path integral,\cite{Mensky}
however, in some sense we consider the classical (not quantum) path 
integral. 
	Let us also mention that the quantum nondemolition 
measurements\cite{Braginsky} are outside the scope of our study, we 
consider only the measurements at the so-called ``standard quantum
limit''.

\begin{figure}
\centerline{
\epsfxsize=3.0in
\epsfbox{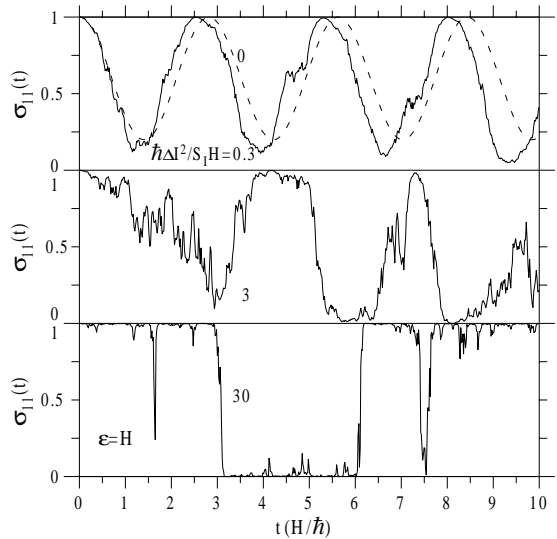}} 
\caption{
Random evolution of $\sigma_{11}$ (particular Monte-Carlo realizations)
for asymmetric double-dot, $\varepsilon =H$, with the electron initially
in the first dot, $\sigma_{11}(0)=1$, for different 
strength of coupling with detector: 
${\cal C}=\hbar (\Delta I)^2/S_IH=$0.3, 3, and 30
from top to bottom. Dashed line represents ${\cal C}=0$ (unmeasured
double-dot). Increasing coupling with detector destroys the quantum
oscillations (while wavefunction remains pure at any $t$), slows down
the transitions between states (Quantum Zeno effect), and for 
${\cal C}\gg 1$ leads to uncorrelated jumps between well localized
states.
 }
\label{fig2}\end{figure}

        Figure \ref{fig2} shows the particular results of the Monte-Carlo
simulations for the double-dot with $\varepsilon =H$ and the different
strength of the interaction with an ideal detector. The electron is initially
located in the first dot, $\sigma_{11}(0)=1$. The dashed line shows
the evolution of $\sigma_{11}$ without detector. Notice that because 
$\varepsilon \neq 0$, the initial asymmetry of the electron location 
remains in this case for infinite time. When the interaction with detector, 
${\cal C} =\hbar(\Delta I)^2/S_IH$, is relatively
small (top solid line), the evolution of $\sigma_{11}$ is close to
that without the detector. However, the electron gradually ``forgets''
the initial asymmetry and the evolution can be described as the slow
variation of the two-parametric phase of oscillations (recall that 
the wavefunction remains pure). In the decoherence approach 
(averaging over realizations) this corresponds to $\sigma_{11}\rightarrow 
1/2$ at $t\rightarrow \infty$. \cite{Gurvitz}

        When the coupling with the detector increases, the 
evolution significantly changes (middle and bottom curves in Fig.\
\ref{fig2}). First, the transition between dots slows down
(Quantum Zeno effect 
\cite{Misra,Carmichael,Plenio,Presilla,Gagen,Itano,Gurvitz,%
Stodolsky,Shnirman}).
Second, while the frequency of transitions decreases with increasing
interaction with the detector,
the time of a transition also decreases,
so eventually we can talk about uncorrelated ``quantum jumps'' 
between states. 

        In a regime of small coupling with a detector, ${\cal C}\ll 1$, 
the detector output is too noisy to follow the
evolution of $\sigma_{ii}$ and, correspondingly, only slightly 
affects the oscillations (the presence of quantum oscillations
in the double-dot can be noticed only as a relatively small
peak in the spectral density of the detector current). In contrast, 
when ${\cal C}\gg 1$ the detector accurately indicates the 
position of electron and simultaneously
destroys the oscillations.  

        Equations (\ref{s11-g})--(\ref{s12-g}) with the term $\cal R$ 
given by Eq.\ (\ref{Rnew}) can be used to obtain  
the evolution of the density matrix in an experiment provided
the detector output $I(t)$ 
and initial condition $\sigma_{ij}(0)$ are known. 
        Notice that even if the initial state is completely
random, $\sigma_{11}=\sigma_{22}=1/2$, $\sigma_{12}=0$, the
nondiagonal matrix element gradually appears during the measurement, 
so that sufficiently long 
observation with an ideal detector leads to almost pure 
wavefunction for the double-dot. Such a purification of the density 
matrix described by Eqs.\ (\ref{s11-g})--(\ref{s12-g}) is analogous to
the localization at $H=0$. 

        Equations (\ref{s11-g})--(\ref{s12-g}) can be generalized
for a nonideal detector, $\Gamma_d > (\Delta I)^2/4S_I$ 
(as in Refs.\ \cite{Stodolsky,Shnirman}), 
which gives less information than possible in principle. 
Let us model it 
as two ideal detectors ``in parallel'' with unaccessible 
output of the second detector.
Then the information loss can be represented
by the extra decoherence term $-\gamma_d \sigma_{12}$ in Eq.\
(\ref{s12-g}) where $\gamma_d=\Gamma_d-(\Delta I)^2/4S_I$.\cite{extraterm} 
The limiting case of a nonideal detector is the detector 
with no output (just an environment, $\Delta I=0$) or with disregarded 
output. Then Eqs.\ (\ref{s11-g})--(\ref{s12-g}) reduce to 
the standard decoherence approach. 

        For a nonideal detector it is meaningful to keep our old 
definition of the localization time, $\tau_{loc}=\tau_{dis}=
2S_I/(\Delta I)^2$, while $\tau_d<2\tau_{loc}$. 
So, we consider localization time not as a real physical quantity  
but as a quantity related to the observer's information. Similarly, 
the effective decoherence time is defined as $\tau_d'=\gamma_d^{-1}$.

\begin{figure}
\centerline{
\epsfxsize=3.0in
\epsfbox{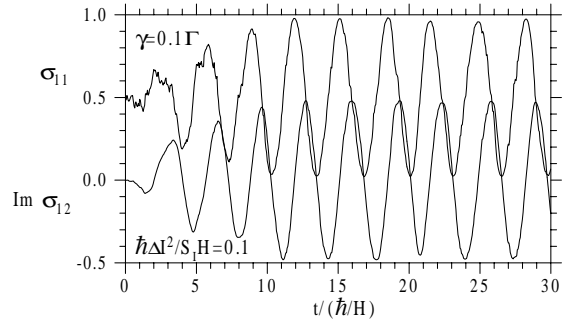}} 
\caption{Gradual purification of the density matrix [$\sigma_{11}(t)$
and $\mbox{Im}\sigma_{12}(t)$ are shown] of the symmetric 
double-dot ($\varepsilon=0$) measured by slightly nonideal 
($\gamma_d/\Gamma_d=0.1$) weakly coupled (${\cal C}=0.1$) detector.
 }
\label{fig3}\end{figure}

        Figure \ref{fig3} shows a particular realization of 
random evolution of $\sigma_{11}$ and $\mbox{Im}\sigma_{12}$ for 
a symmetric double dot measured by weakly coupled (${\cal C}=0.1$) nonideal 
detector with $\gamma_d /\Gamma_d=0.1$. We start from maximally mixed 
state, $\sigma_{11}(0)=0.5$, $\sigma_{12}(0)=0$, and the Figure shows  
the gradual purification of the density matrix in a course of measurement
(notice that $\mbox{Re}\sigma_{12}(t)=0$ because $\varepsilon =0$). 
The nonideality of the detector does not allow the complete purification:
oscillations of $\mbox{Im}\sigma_{12}(t)$ do not reach $\pm 0.5$ limit, 
as it would be in the case of ideal detector. 

        Let us mention that following the ``orthodox'' (Copenhagen) 
point of view, we do not attempt to distinguish between 
``real'' density matrix and the density matrix which can be known by 
the observer.
For example, the evolution of $\sigma_{11}$ due to the measurement 
in case of no tunneling between dots ($H=0$) can be interpreted both 
as a real process or just as gradual acquiring of information about
the electron position. Another example is the case of nonideal detector. 
We can interpret the term $-\gamma_d \sigma_{12}$ in Eq.\ (\ref{s12-g})
as real decoherence, however, it is also possible to argue that it 
just represents the partial loss of information inside the imperfect 
detector, so that perhaps the pure density matrix could be restored 
if some hidden traces left in the detector had been analyzed. Developing 
this example further, let us imagine that two 
observers have different levels of access to the detector information,
then the density matrix for them will be different. Actually, this
just means that the observer with less information will not be able
to make as many (or as accurate) predictions as the other one. 
Nevertheless, he still can treat his density matrix as a real one for
all purposes. 
        The limiting case when the observer does not have any information
about the detector output (or this information is ignored in 
the experiment) is equivalent to averaging over all possible realizations,
i.e.\ to the standard decoherence approach. 

        So, if different realizations of the detector output are 
effectively averaged in an experiment (as in Ref.\ \cite{Buks}), 
the decoherence approach is suitable. In contrast, if the single 
realization of the detector current is recorded (and somehow used) in 
an experiment, then the proper description 
is given by Eqs.\ (\ref{s11-g})--(\ref{s12-g}). 
The simplest experimental idea is just to measure $I(t)$ when 
$\cal C$ is not too small and check if it is consistent
with these equations. However, it would be much more interesting 
to devise an experiment in which the subsequent system 
evolution depends on the preceding measurement result.

        For example, let us first prepare the double-dot in
the symmetric coherent state, $\sigma_{11}=\sigma_{22}=|\sigma_{12}|=1/2$, 
make $H=0$ (raise the barrier), and begin measurement 
with an almost ideal detector.  
According to our formalism, after some time $\tau$ 
(the most interesting case is $\tau \sim \tau_{loc}$) 
the wavefunction remains pure but becomes asymmetric and 
can be calculated with Eqs.\ (\ref{s1-2}) and (\ref{s_12-g}). 
To prove this, an experimentalist can use the knowledge of the 
wavefunction to move the electron into the first dot with
probability equal to unity. Namely, he switches off 
the detector at $t=\tau $, reduces the barrier (to create finite $H$), 
and creates the energy difference 
$\varepsilon = [(1-4|\sigma_{12}|^2)^{1/2}-1] 
H \mbox{Re}\sigma_{12}/ |\sigma_{12}|^{2}$; 
then after the time period
$\Delta t =[\pi-\arcsin (\mbox{Im} \sigma_{12} \, \hbar\Omega/H)]/\Omega$
the ``whole'' electron will be moved to the first dot, 
that can be checked by the detector switched on again. 
(Alternatively, using the knowledge of $\sigma_{ij}(\tau )$ 
an experimentalist can produce exactly the ground state of the 
double-dot system and check it, for example, by photon
absorption.) 

        Another experimental idea is to demonstrate 
the gradual purification of the double-dot density matrix. 
Let us start with a completely random state 
($\sigma_{11}=\sigma_{22}=1/2$, $\sigma_{12}=0$) 
of the double-dot with finite $H$.  
Then using the detector output $I(t)$ and Eqs.\ 
(\ref{s11-g})--(\ref{s12-g}) it is possible to calculate
the evolution of the density matrix. These calculations
will show the gradual purification 
(the most interesting case is $\Omega \tau_{loc}\lesssim 1$),
eventually ending up with almost pure wavefunction with 
precisely known {\it phase} of quantum oscillations. 
The final check of the wavefunction can be similar to that 
considered above. However, it can be even simpler, 
because with the knowledge of the phase of oscillations
it is easy to stop the evolution by raising the barrier when the 
electron is with certainty in the first dot. If rapid calculations 
(by some analog on-chip circuit) are not available, the barrier control 
can be random, while appropriate cases can be selected later.  

        An experiment of this kind could verify the formalism 
developed in the present paper. While such an experiment is
still a challenge for present-day technology, we 
hope that it can be realized in the near future.

        In conclusion, we have developed a simple formalism 
for the evolution of the double-dot density matrix with an account
of the result of the continuous measurement by weakly
coupled (weakly responding) point contact. In contrast to most
previous studies on the selective quantum measurements, our
equations treat mixed states and allow the consideration of
a nonideal detector. The equations show the gradual purification
of initially mixed state of the double-dot due to continuous
quantum measurement. This effect can be studied experimentally 
in various mesoscopic setups. 

        The author thanks S. A. Gurvitz, K. K. Likharev, 
D. V. Averin, and T. V. Filippov 
for fruitful discussions. The work was supported in 
part by French MENRT (PAST), US AFOSR, and Russian RFBR.


\begin{references}
\bibitem{bigDI}  We implicitly assume that the corresponding detector 
	collapse happens on the time scale $e/I_0 \ll t_c\ll 
	(e/I_0)(I_0/\Delta I)^2$, much faster than the typical evolution 
	of the double-dot density matrix.
	In the case $\Delta I \sim I_0$ the evolution would strongly 
        depend on the detector interaction with  
        the next measuring stage.
\bibitem{largeT} In the case $1-{\cal T}\ll 1$, Eq.\ (\protect\ref{weak})
        should be replaced by $|\Delta I| \ll (1-{\cal T}) I_0
        \sim S_I/e$.
\bibitem{phase} Modification of Eq.\ (\protect\ref{gam_d-m}) in the case when 
        the phase of transmitted and reflected electrons in the detector 
        is sensitive to the double-dot state has been discussed in
        Ref.\ \protect\cite{Stodolsky}. We assume that there is no such
        a dependence in our case.
\bibitem{Bayes} A posteriori probability $\tilde{p}(A)$ of an event 
        $A$ given that the event $F$ has happened, 
        is equal to $\tilde{p}(A)=p(A)p(F|A)/\sum_B [p(B)p(F|B)]$ where 
        $p(A)$ is a priori probability and $p(F|A)$ is the conditional 
        probability of event $F$ given event $A$.
\bibitem{strong} If a similar approach is formally applied to 
        the case  $\Delta I \sim I_0$, we obtain  
        $|\langle \sigma_{12}(\tau )\rangle_r | /[\sigma_{11}(0)
        \sigma_{22}(0)]^{1/2} \leq [2\sqrt{S_{I1} S_{I2}}/
        (S_{I1}+S_{I2})]^{1/2} \exp [-\tau (\Delta I)^2 /
        2(S_{I1}+S_{I2})] $. Then the corresponding result of 
        Ref.\ \cite{Gurvitz} 
        is smaller than this upper bound for large $\tau $, while 
        it is above the limit for $\tau \alt \Gamma_d^{-1} \sim e/I_0$. 
        The latter unphysical
        situation is because the detector cannot be described classically
        at small $\tau$.
\bibitem{alternat} Notice that Eqs.\ (\protect\ref{s1-2}) and 
        (\protect\ref{s_12-g}) (for $\varepsilon =0$) can be readily 
        expressed using the 
        standard ``reduction'' procedure, $\sigma (\tau ) = 
        {\cal A}/\mbox{Tr} {\cal A}$, ${\cal A} =
        {\cal P} (\langle I\rangle ,\tau ) \, \sigma (0) \, 
        {\cal P} (\langle I\rangle ,\tau )$, if the generalized
        ``projection'' operator ${\cal P} (\langle I\rangle ,\tau )$
        is defined as ${\cal P}_{ii}=[P_i(\langle I\rangle ,\tau )]^{1/2}$,
        ${\cal P}_{12}={\cal P}_{21}=0$.
\bibitem{extraterm} If the level asymetry $\varepsilon$ is directly 
	influenced by the detector state (as in the case of single-electron
	transistor\protect\cite{Shnirman}) and this influence is
	different for currents $I_1$ and $I_2$ through the detector,
	then the corresponding compensating term should be added into Eq.\
	(\ref{s12-g}) -- see A. N. Korotkov, cond-mat/9906439.
\end{references}
\end{document}